\documentclass[longauth]{aa}
\makeatletter
\renewcommand*\aa@pageof{, page \thepage{} of \pageref*{LastPage}}

\usepackage{graphicx}
\usepackage{txfonts}
\usepackage{natbib}
\bibpunct{(}{)}{;}{a}{}{,}
\usepackage[breaklinks, colorlinks, citecolor=blue, linkcolor=blue]{hyperref}

\usepackage[autostyle,english=british]{csquotes}

\newcommand{\rjup}{\ensuremath{R_\mathrm{Jup}}\xspace}

\newcommand{\ktrue}{\ensuremath{k_\mathrm{true}}\xspace}

\newcommand{\teff}{\ensuremath{T_\mathrm{eff}}\xspace}

\usepackage{orcidlink}
\begin{document} 

\title{Validation of up to seven TESS planet candidates through multi-colour transit photometry using MuSCAT2 data}\titlerunning{TOI validation with MuSCAT2}\authorrunning{A. Peláez-Torres et al.} 

\author{A.~Pel\'aez-Torres\inst{\ref{IAC}}\fnmsep\inst{\ref{ull}}\fnmsep\inst{\ref{IAA}}\orcidlink{0000-0001-9204-8498}
    \and E.~Esparza-Borges\inst{\ref{IAC}}\fnmsep\inst{\ref{ull}}\orcidlink{0000-0002-2341-3233} \and
    E.~Pall\'e\inst{\ref{IAC}}\fnmsep\inst{\ref{ull}}\orcidlink{0000-0003-0987-1593} \and
    H.~Parviainen\inst{\ref{ull}}\fnmsep\inst{\ref{IAC}}\orcidlink{0000-0001-5519-1391} \and F.~Murgas\inst{\ref{IAC}}\fnmsep\inst{\ref{ull}}\orcidlink{0000-0001-9087-1245} \and G.~Morello\inst{\ref{IAA}}\fnmsep\inst{\ref{IAC}}\orcidlink{0000-0002-4262-5661} \and M.R.~Zapatero-Osorio\inst{\ref{astrobio}}\orcidlink{0000-0001-5664-2852} \and J.~Korth\inst{\ref{lund}}\orcidlink{0000-0002-0076-6239} \and N.~Narita\inst{\ref{Komaba}}\fnmsep\inst{\ref{ABC}}\fnmsep\inst{\ref{IAC}}\orcidlink{0000-0001-8511-2981} \and A.~Fukui\inst{\ref{Komaba}}\fnmsep\inst{\ref{IAC}}\orcidlink{0000-0002-4909-5763} \and I.~Carleo\inst{\ref{IAC}}\fnmsep\inst{\ref{INAF_torino}}\orcidlink{0000-0002-0810-3747} 
    \and R.~Luque\inst{\ref{chicago}}\orcidlink{0000-0002-4671-2957} \and 
    N.~Abreu García\inst{\ref{IAC}}\fnmsep\inst{\ref{ull}}\orcidlink{0009-0002-5067-5463} \and
    K.~Barkaoui\inst{\ref{belg}}\fnmsep\inst{\ref{mit}}\fnmsep\inst{\ref{IAC}}\orcidlink{0000-0003-1464-9276} \and A.~Boyle\inst{\ref{california}}\orcidlink{0000-0001-6037-2971} \and V. J. S.~Béjar\inst{\ref{IAC}}\fnmsep\inst{\ref{ull}}\orcidlink{0000-0002-5086-4232} \and 
    Y.~Calatayud-Borras\inst{\ref{IAC}}\fnmsep\inst{\ref{ull}} \and D.V.~Cheryasov\inst{\ref{sternberg}} \and J.L.~Christiansen \inst{\ref{california}}\orcidlink{0000-0002-8035-4778} \and D.R.~Ciardi\inst{\ref{california}}\orcidlink{0000-0002-5741-3047} 
    \and G.~Enoc\inst{\ref{IAC}}\fnmsep\inst{\ref{ull}}\orcidlink{0000-0003-0597-7809} \and Z.~Essack\inst{\ref{mit}}\fnmsep\inst{\ref{kavli}}\orcidlink{0000-0002-2482-0180} \and I.~Fukuda\inst{\ref{dep_multi_tokyo}}\orcidlink{0000-0002-9436-2891} \and
    G.~Furesz\inst{\ref{kavli}}\orcidlink{0000-0001-8467-9767} \and D.~Galán\inst{\ref{IAC}}\fnmsep\inst{\ref{ull}}\orcidlink{0000-0001-6191-8251} \and S.~Geraldía-González\inst{\ref{IAC}}\fnmsep\inst{\ref{ull}} \and
    S. Giacalone\inst{\ref{berkeley}}\orcidlink{0000-0002-8965-3969}\and
    H.~Gill\inst{\ref{berkeley}}\orcidlink{0000-0001-6171-7951} \and E. J.~Gonzales\inst{\ref{santacruz}}\orcidlink{0000-0002-9329-2190} \and Y. ~Hayashi\inst{\ref{dep_multi_tokyo}}\orcidlink{0000-0001-8877-0242} \and K.~Ikuta\inst{\ref{dep_multi_tokyo}}\orcidlink{0000-0002-5978-057X} \and K. ~Isogai\inst{\ref{dep_multi_tokyo}}\orcidlink{0000-0002-6480-3799} 
    \and T.~Kagetani\inst{\ref{dep_multi_tokyo}}\orcidlink{0000-0002-5331-6637} \and 
    Y.~Kawai \inst{\ref{dep_multi_tokyo}}\orcidlink{0000-0002-0488-6297} \and    
    K.~Kawauchi\inst{\ref{dept_kusatsu}}\fnmsep\inst{\ref{dep_multi_tokyo}}\orcidlink{0000-0003-1205-5108} \and 
    P. ~Klagyvik\inst{\ref{germ}} \and 
    T. ~Kodama\orcidlink{0000-0001-9032-5826} \and 
    N.~Kusakabe\inst{\ref{ABC}}\fnmsep\inst{\ref{naoj}}\orcidlink{0000-0001-9194-1268} \and 
    A.~Laza-Ramos\inst{\ref{valencia}}\orcidlink{0000-0003-3316-3044} \and 
    J.P.~de Leon\inst{\ref{dep_multi_tokyo}}\orcidlink{0000-0002-6424-3410} \and
    J.~H.~Livingston\inst{\ref{ABC}}\fnmsep\inst{\ref{naoj}}\fnmsep\inst{\ref{dept_mitaka}}\orcidlink{0000-0002-4881-3620} \and M.~B.~Lund\inst{\ref{california}}\orcidlink{0000-0003-2527-1598} \and A.~Madrigal-Aguado\inst{\ref{IAC}}\orcidlink{0000-0002-9510-0893} \and P.~Meni\inst{\ref{IAC}}\orcidlink{0009-0001-7943-0075} \and M.~Mori\inst{\ref{dep_multi_tokyo}}\orcidlink{0000-0003-1368-6593} \and S.~Muñoz Torres\inst{\ref{IAC}}\fnmsep\inst{\ref{ull}}\orcidlink{0000-0003-4269-4779} \and 
    J.~Orell-Miquel\inst{\ref{IAC}}\fnmsep\inst{\ref{ull}}\orcidlink{0000-0003-2066-8959} \and 
    M.~Puig\inst{\ref{IAA}}\orcidlink{} \and G.~Ricker\inst{\ref{mit}}\orcidlink{0000-0003-2058-6662} \and 
    M.~Sánchez-Benavente\inst{\ref{IAC}}\fnmsep\inst{\ref{ull}}\orcidlink{0000-0003-2693-279X} \and 
    A.B.~Savel \inst{\ref{maryland}}\orcidlink{0000-0002-2454-768X} \and J.E.~Schlieder\inst{\ref{greenbelt}} \and R.P.~Schwarz\inst{\ref{CfA}}\orcidlink{0000-0001-8227-1020} \and 
    R.~Sefako\inst{\ref{capetown}}\orcidlink{0000-0003-3904-6754} \and 
    P.~Sosa-Guillén\inst{\ref{IAC}}\fnmsep\inst{\ref{ull}} \and 
    M.~Stangret\inst{\ref{INAF_padova}}\orcidlink{0000-0002-1812-8024} \and C.~Stockdale\inst{\ref{hazelwood}}\orcidlink{0000-0003-2163-1437} \and M.~Tamura\inst{\ref{dep_tokyo}}\fnmsep\inst{\ref{ABC}}\fnmsep\inst{\ref{naoj}}\orcidlink{0000-0002-6510-0681} \and Y.~Terada\orcidlink{0000-0003-2887-6381} \and J.D.~Twicken\inst{\ref{seti}}\orcidlink{0000-0002-6778-7552} \and 
    N.~Watanabe\inst{\ref{dep_multi_tokyo}}\orcidlink{0000-0002-7522-8195} \and J.~Winn\inst{\ref{princeton}}\orcidlink{0000-0002-4265-047X} \and S.G.~Zheltoukhov\inst{\ref{sternberg}}\orcidlink{0000-0002-6321-0924} \and C.~Ziegler\inst{\ref{austin}} \and
    Y.~Zou}

    \institute{Instituto de Astrof\'isica de Canarias (IAC), E-38200 La Laguna, Tenerife, Spain\label{IAC}
         \and
         Departamento de Astrof\'isica, Universidad de La Laguna (ULL), E-38206 La Laguna, Tenerife, Spain\label{ull}
         \and Instituto de Astrofísica de Andalucía (IAA-CSIC), Gta. de la Astronomía s/n, 18008 Granada, Granada, Spain
         \label{IAA} \and
         Lund Observatory, Division of Astrophysics, Department of Physics, Lund University, Box 118, 22100 Lund, Sweden \label{lund}
         \and
         Komaba Institute for Science, The University of Tokyo, 3-8-1 Komaba, Meguro, Tokyo 153-8902, Japan\label{Komaba}
         \and
         Astrobiology Center, 2-21-1 Osawa, Mitaka, Tokyo 181-8588, Japan\label{ABC}
         \and
         Center for Astrophysics \textbar \ Harvard \& Smithsonian, 60 Garden Street, Cambridge, MA 02138, USA \label{CfA}
         \and
         INAF -- Osservatorio Astrofisico di Torino, Via Osservatorio 20, I-10025, Pino Torinese, Italy\label{INAF_torino}
         \and
         Department of Space, Earth and Environment, Chalmers University of Technology, SE-412 96 Gothenburg, Sweden\label{chalmers}
         \and
         Department of Astronomy, Graduate School of Science, The University of Tokyo, 7-3-1 Hongo, Bunkyo-ku, Tokyo 113-0033,Japan\label{dep_tokyo}
         \and
         National Astronomical Observatory of Japan, 2-21-1 Osawa,Mitaka, Tokyo 181-8588, Japan\label{naoj}
         \and
         Department of Astronomy \& Astrophysics, University of Chicago, Chicago, IL 60637, USA\label{chicago}
         \and
         INAF – Osservatorio Astronomico di Padova, Vicolo dell’Osservatorio 5, 35122, Padova, Italy\label{INAF_padova}
         \and
         Department of Multi-Disciplinary Sciences, Graduate School of Arts and Sciences, The University of Tokyo, 3-8-1 Komaba, Meguro, Tokyo 153-8902, Japan\label{dep_multi_tokyo}
         \and
         Department of Astronomical Science, The Graduated University for Advanced Studies, SOKENDAI, 2-21-1, Osawa, Mitaka, Tokyo, 181-8588, Japan\label{dept_mitaka}
         \and
         Department of Physical Sciences, Ritsumeikan University, Kusatsu, Shiga 525-8577, Japan\label{dept_kusatsu}
         \and
         Departamento de Astronomía y Astrofísica, Universidad de Valencia (UV), E-46100, Burjassot, Valencia, Spain\label{valencia}
         \and
         American Association of Variable Star Observers, 185 Alewife Brook Parkway, Suite 410, Cambridge, MA 02138, USA\label{american}
         \and
         Hazelwood Observatory, Australia\label{hazelwood}
         \and
         South African Astronomical Observatory, P.O. Box 9, Observatory, Cape Town 7935, South Africa\label{capetown}
         \and
         Sternberg Astronomical Institute, M.V. Lomonosov Moscow State University, 13, Universitetskij pr., 119234, Moscow, Russia \label{sternberg}
         \and
         Department of Astronomy, University of California Berkeley, Berkeley, CA 94720, USA
         \label{berkeley}
         \and
         Department of Astronomy, University of Maryland, College Park, College Park, MD 20742 USA
         \label{maryland}
         \and
         NASA Exoplanet Science Institute – Caltech/IPAC 1200 E. California Blvd Pasadena, CA 91125 USA
         \label{california}
         \and
         Department of Astronomy and Astrophysics, University of California, Santa Cruz, 1156 High St. Santa Cruz , CA 95064, USA
         \label{santacruz}
         \and
         Exoplanets and Stellar Astrophysics Laboratory, NASA Goddard Space Flight Center, Greenbelt, MD, USA
         \label{greenbelt}
         \and
         SETI Institute, Mountain View, CA 94043 USA/NASA Ames Research Center, Moffett Field, CA 94035 USA
         \label{seti}
         \and
         Department of Earth, Atmospheric and Planetary Sciences, Massachusetts Institute of Technology, Cambridge, MA 02139, USA
         \label{mit}
         \and
         Department of Physics and Kavli Institute for Astrophysics and Space Research, Massachusetts Institute of Technology, Cambridge, MA 02139, USA
         \label{kavli}
         \and
         Astrobiology Research Unit, Universit\'e de Li\`ege, All\'ee du 6 Ao\^ut 19C, B-4000 Li\`ege, Belgium
         \label{belg}
         \and
         Department of Astrophysical Sciences, Princeton University, Princeton, NJ 08544, USA\label{princeton}
         \and
         Department of Physics, Engineering and Astronomy, Stephen F. Austin State University, 1936 North St, Nacogdoches, TX 75962, USA
         \label{austin}
         \and
         Centro de Astrobiologia (CSIC-INTA), Carretera de Ajalvir km 4, 28850 Torrejon de Ardoz, Madrid, Spain
         \label{astrobio}
         \and
         Freie Universit\"at Berlin, Institute of Geological Sciences, Malteserstr. 74-100, 12249 Berlin, Germany
         \label{germ}}

        \date{Received month day, year; accepted month day, year}

\abstract
{The TESS mission searches for transiting exoplanets by monitoring the brightness of hundreds of thousands of stars across the entire sky. M-type planet hosts are ideal targets for this mission due to their smaller size and cooler temperatures, which makes it easier to detect smaller planets near or within their habitable zones. Additionally, M~dwarfs have a smaller contrast ratio between the planet and the star, making it easier to measure the planet's properties accurately. 
Here, we report the validation analysis of 13 TESS exoplanet candidates orbiting around M dwarfs. We studied the nature of these candidates through a multi-colour transit photometry transit analysis using several ground-based instruments (MuSCAT2, MuSCAT3, and LCO-SINISTRO), high-spatial resolution observations, and TESS light curves. We present the validation of five new planetary systems: TOI-1883b, TOI-2274b, TOI2768b, TOI-4438b, and TOI-5319b, along with compelling evidence of a planetary nature for TOIs 2781b and 5486b. We also present an empirical definition for the Neptune desert boundaries. The remaining six systems could not be validated due to large true radius values  overlapping with the brown dwarf regime or, alternatively, the presence of chromaticity in the MuSCAT2 light curves.}


\maketitle
%

\section{Introduction}

Since it was launched on 18 April 2018, the Transiting Exoplanet Survey Satellite  mission \citep[TESS, ][]{ricker2015} has provided 346 validated exoplanets and more than 6,599 planet candidates\footnote{From the \texttt{NASA Exoplanet Exploration (TESS)}}. However, not every planet-like TESS object of interest (TOI) ends up being a true exoplanet. There are many other possible astronomical scenarios that can mimic a planetary signal, such as the presence of a brown dwarf or a background binary system. 

The nature of the host star plays a key role in determining the appropriate method for validating an exoplanet. Thus, for those planets hosted by stars too faint for performing radial velocity studies, it is extremely helpful to observe and analyse multi-colour photometry from ground-based telescopes and space missions to evaluate possible background contaminants, allowing us to distinguish whether the source of the transit signal is a planet or not. If we were to study an inactive, bright, and slowly rotating host star, we could validate whether the candidate is a planet by obtaining its mass through radial velocity measurements. However, considering the characteristics of our targets, most planets around M dwarfs are too faint for precise radial velocity measurements, ground-based multi-colour transit photometry is the best method to validate the nature of these candidates. By analysing the simultaneous light curves in several filters, we evaluated the possible brightness contamination of the target by an unresolved source and we were able to estimate the 'true' planet candidate radius through the ratio between the estimated stellar radius and the uncontaminated radius \citep{drake2003, tingley2004, parviainen2019, 
parviainen2020, parviainen2021, esparza2022, fukui2022, morello2023, murgas2022}.

With the discovery of more than 5,000 exoplanets, the relation between the radii and the orbital period validated the existence of a significant dearth of Neptune-sized ($\sim3-4 R_\oplus$) planets in orbital periods between 2-4 days. This scarcity of objects cannot be explained by observational biases, as other Neptune-sized planets have been found with longer periods and it is easier to detect planets with shorter periods mostly by using either transits or radial velocities methods. This region has been called the 'Neptune desert' or 'photoevaporation desert' \citep{sanchis2014, owen2017, jenkins2020}. The photoevaporation phenomenon is the loss of atmospheric mass due to high-energy irradiation from the host star. Although the number of known exoplanets in the desert is low, they have a huge scientific value since these planets are either halfway to losing their atmosphere or still have it \citep{lundkvist2016, mazeh2016, lopez2017}. 

In this work, we analyse multi-colour photometric observations of 13 TESS planet candidates orbiting faint M dwarf hosts: TOI-1883.01, TOI-2274.01, TOI-2603.01, TOI-2768.01, TOI-2781.01, TOI 4438.01, TOI-5205.01, TOI-5268.01, TOI-5319.01, TOI-5344.01, TOI-5464.01, TOI-5486.01, and TOI-5641.01. We used ground-based photometric observations in \textit{g}, \textit{r}, \textit{i}, and \textit{$z_s$} bands from MuSCAT2, MuSCAT3, and 1m-LCO photometry together with TESS photometry.

In Sections~\ref{sec: TESS_Photometry},~\ref{sec: Ground_based_observations}, and~\ref{sec: Stellar_Parameters}, we describe the observations used in this study and the stellar parameters for each candidate. The data treatment and the methods applied (light curve, validation and contamination analysis) are explained in Section \ref{sec: Methods}. We present our results in Section~\ref{sec: Results}. Finally, we discuss the planetary nature of our candidates and conclude our analysis in Section~\ref{sec: Discussion}.

\section{TESS photometry}
\label{sec: TESS_Photometry}

TESS observations involve high-cadence data collection in specific sectors of the sky. The collected data were processed at the TESS Science Processing Operations Center (SPOC; \cite{jenkins2016}). A transit search was then carried out with the combined Presearch Data Conditioning Simple Aperture Photometry (PDCSAP; \cite{stumpe2012, stumpe2014, smith2012}) light curves with an adaptive, noise-compensating matched filter \citep{jenkins2002, jenkins2010, jenkins2020}. This produced a threshold-crossing event (TCE) with specific orbital periods for which an initial limb-darkened transit model was fitted \citep{li2019} and a suite of diagnostic tests were conducted to help assess the possible planetary nature of the signal \citep{twicken2018}. The TESS Science Office reviews and evaluates the data, occasionally issuing alerts. This detection process applies to TOI-2274.01, TOI-2603.01, and TOI-4438.01, which had 2 min cadence observations.

The rest of the TOIs in our sample were first detected in the inspection of the full-frame image data using the quick-look pipeline (QLP; \cite{huang2020, kunimoto2021, kunimoto2022}) and the faint target transit search at MIT. We summarise the TESS observations of all our targets in Table \ref{tab:targets_photometry_date}, including the TESS Input Catalog (TIC) number, TESS Sector(s), and the signal-to-noise ratio (S/N) of the TESS observation.

\subsection{Sample selection}


The MUSCAT2 instrument \citep{narita2019} at the Carlos Sanchez Telescope in the Teide Observatory is dedicated to follow-ups of TESS planet candidates. Among its various follow-up activities, MUSCAT2 is especially well suited for the validation of planets around faint M dwarf stars. We prioritised these observations at a high photometric accuracy to validate their planetary nature via multi-colour observations (see the section on methods). In this work, we summarise our observational efforts during the 2021 and 2022 campaigns, when we gathered successful observations for 13 M dwarf planet candidates suitable for multi-colour photometry validation.
After our sample was defined, we searched for and downloaded all the available data on these same objects obtained within the TESS Followup Official Program (TFOP) and available via ExoFOP\footnote{\url{https://exofop.ipac.caltech.edu/tess/}}.

\begin{table*}[ht]
\caption{Targets, TICs, TESS sectors, S/N, instruments used for observations, and filters and dates for each target in the sample.}
\centering
\begin{tabular}{lllllll}
\hline\hline\\
TOI ID & TIC & TESS Sector & TESS S/N & \begin{tabular}[c]{@{}l@{}}Ground-Based\\ Photometry\end{tabular} & Filters & Date \\[14pt] \hline\\[0.01cm]

1883.01 & 348755728.01 & 8, 35 & 11 & \begin{tabular}[c]{@{}l@{}}MuSCAT2,\\ MuSCAT3\end{tabular} & \begin{tabular}[c]{@{}l@{}}\textit{g}, \textit{r}, \textit{i}, \textit{$z_s$}\\\textit{g}, \textit{r}, \textit{i}, \textit{$z_s$}\end{tabular} & \begin{tabular}[c]{@{}l@{}}14.03.2021,\\ 23.03.2022,\\ 11.02.2021\end{tabular}  \\[15pt]

2274.01 & 289164482.01 & 14, 26, 40 & 10.7 & \begin{tabular}[c]{@{}l@{}}MuSCAT2,\\ MuSCAT3,\\ SINISTRO,\\ SINISTRO\end{tabular} &   \begin{tabular}[c]{@{}l@{}}\textit{g}, \textit{r}, \textit{i}, \textit{$z_s$}\\\textit{g}, \textit{r}, \textit{i}, \textit{$z_s$}\\ \textit{$z_s$}\\ \textit{$z_s$}\end{tabular} & \begin{tabular}[c]{@{}l@{}}19.03.2021,\\ 19.03.2022,\\ 21.10.2020,\\ 09.08.2021\end{tabular}  \\[25pt]

2603.01 & 176772671.01 & 35 & 24.9 & MuSCAT2 & \textit{g}, \textit{r}, \textit{i}, \textit{$z_s$} & 30.12.2021 \\[15pt]

2768.01 & 43556801.01 & 6, 33 & 15 & \begin{tabular}[c]{@{}l@{}}MuSCAT2,\\ SINISTRO\end{tabular} & \begin{tabular}[c]{@{}l@{}}\textit{g}, \textit{r}, \textit{i}, \textit{$z_s$}\\ \textit{i}\end{tabular} & \begin{tabular}[c]{@{}l@{}}18.10.2021,\\ 23.01.2022\end{tabular} \\[15pt]

2781.01 & 317417916.01 & 6, 33 & 17 & MuSCAT2 & \textit{g}, \textit{r}, \textit{i}, \textit{$z_s$} & 08.11.2021 \\[15pt]

4438.01 & 22233480.01 & 40 & 14.2 & MuSCAT2 & \textit{g}, \textit{r}, \textit{i}, \textit{$z_s$} & \begin{tabular}[c]{@{}l@{}}16.05.2022,\\ 01.05.2023\end{tabular} \\[15pt]

5205.01 & 419411415.01 & 41 &  & MuSCAT2 & \textit{g}, \textit{r}, \textit{$z_s$} & 20.05.2022 \\[25pt]

5268.01 & 202468443.01 & 41, 48, 49 & 24 & MuSCAT2 & \textit{g}, \textit{r}, \textit{i}, \textit{$z_s$} & 24.03.2022 \\[15pt]

5319.01 & 246965431.01 & 18, 42, 43 & - & MuSCAT2 & \textit{g}, \textit{r}, \textit{i}, \textit{$z_s$} &   20.11.2022   \\[15pt]

5344.01 & 16005254.01 & 43, 44 & 30 & MuSCAT2 & \textit{g}, \textit{r}, \textit{i}, \textit{$z_s$} & 18.12.2022 \\[15pt]

5464.01 & 171646471.01 & 45, 48 & 20 &  \begin{tabular}[c]{@{}l@{}}MuSCAT2,\\ MuSCAT3\end{tabular} &   \begin{tabular}[c]{@{}l@{}}\textit{g}, \textit{r}, \textit{i}, \textit{$z_s$}\\\textit{$z_s$}\end{tabular} & \begin{tabular}[c]{@{}l@{}}18.12.2022,\\ 26.04.2022\end{tabular}  \\[25pt]

5486.01 & 291109653.01 & 23, 46, 50 & 20 & MuSCAT2 & \textit{g}, \textit{r}, \textit{i}, \textit{$z_s$} & \begin{tabular}[c]{@{}l@{}}27.05.2022,\\ 12.02.2023\end{tabular}\\[15pt]

5641.01 & 141202786.01  & 49 & - & MuSCAT2 & \textit{g}, \textit{r}, \textit{i}, \textit{$z_s$} & 08.03.2023 \\[15pt]
\hline\\[0.01cm]
\end{tabular}
\label{tab:targets_photometry_date}
\end{table*}

\section{Ground-based observations}
\label{sec: Ground_based_observations}

\begin{table}[hpbt]
\caption[]{NOT ALFOSC spectroscopic observations.}
\label{tab:alfosc}
\centering
\begin{tabular}{lcccc}
  \hline
  \noalign{\smallskip}
  Target         &  UT date   & Texp & Air mass & SpT \\
                 &            & (s)  &          &     \\
  \noalign{\smallskip}
  \hline
  \noalign{\smallskip}

  TOI-2274 & 2023 Mar 19 & 2$\times$700 & 1.15 & M0.5\,$\pm$\,0.5 \\
  TOI-2603 & 2022 Dec 01 & 2$\times$900 & 1.75 & M1.0\,$\pm$\,0.5 \\  
  TOI-2768 & 2022 Dec 01 & 2$\times$600 & 1.40 & M0.0\,$\pm$\,0.5 \\
  TOI-2781 & 2022 Nov 29 & 2$\times$800 & 1.90 & M1.0\,$\pm$\,0.5 \\  
  TOI-4438 & 2023 Mar 19 & 2$\times$900 & 1.16 & M4.0\,$\pm$\,0.5 \\ 
  TOI-5319 & 2022 Nov 29 & 2$\times$400 & 1.50 & M3.0\,$\pm$\,0.5 \\  
  TOI-5486 & 2022 Nov 30 & 2$\times$600 & 1.70 & M2.5\,$\pm$\,0.5 \\
  \noalign{\smallskip}
  \hline
\end{tabular}
\end{table}

The TESS pixel scale is $\sim 21\arcsec$ pixel$^{-1}$ and photometric apertures typically extend out to roughly 1 arcminute, generally causing multiple stars to become blended within the \textit{TESS} aperture. To try to  determine the true source of the detections in the \textit{TESS} data and refine their ephemerides and transit shapes, we conducted ground-based photometric follow-up observations (also detailed in Table \ref{tab:targets_photometry_date}) as part of the {\tt TESS} Follow-up Observing Program\footnote{\url{https://tess.mit.edu/followup}} Sub Group 1 \citep[TFOP;][]{collins2019}, using several facilities.

\subsection{Optical spectroscopy}

Low-resolution spectra were acquired with the Alhambra Faint Object Spectrograph and Camera (ALFOSC) mounted at the Cassegrain focus of the Nordic Optical Telescope (NOT) on the Roque de los Muchachos Observatory (La Palma island). ALFOSC is equipped with a Teledyne e2v 2048$\times$2048 detector that has a projected pixel size of 0\farcs21 on the sky. During the observations, we windowed the detector along the direction perpendicular to the spectral axis to a size of 500$\times$2048, which is sufficient for the low-resolution, long-slit observing mode. The observing set-up consisted of a 1 arcsec-width slit and the grism \#5, which has a inbuilt order-blocker filter. This instrumental configuration yielded optical spectra with a dispersion of 3.5 \AA\,pix$^{-1}$ over the wavelength interval 5100--9750 \AA~and a spectral resolution of 16 \AA. Table~\ref{tab:alfosc} provides the log of the ALFOSC observations including target names, observing dates (universal time, UT), exposure time, and airmass.

Raw data were corrected for bias and flat-fielded by using well-illuminated images taken before the observing nights. Spectra were optimally extracted with IRAF routines and calibrated in wavelength with ThAr arcs taken with the same instrumental configuration as the targets. Wavelength calibration had an accuracy of 0.4--0.5 \AA. The observations of two spectroscopic flux-standard stars (BD$+$17\,4708 and HD\,84937) on two different nights allowed us to correct the observations for instrumental response. These two sdF standards have tabulated fluxes in IRAF directories. We did not correct for atmospheric extinction due to the Earth's atmosphere because our data mostly covered red optical wavelengths, at which differential extinction is not significant. The two observed spectra per target were combined into a single 1D spectrum to increase the S/N of the data. The final spectra are shown in Figure G.1\footnote{Appendices are available in Zenodo as camera-ready material \url{https://zenodo.org/records/13326858}}, where some of the strongest atomic and molecular features are labeled.

We derived the spectral types of our targets by direct comparison of the ALFOSC data with the library of empirical stellar spectra of \citet{kesseli2017}. The templates cover spectral types O5 through L3 with a resolving power of about 2000. We degraded the resolution of the templates to that of the ALFOSC data. The results are summarised in the column `SpT' of Table~\ref{tab:alfosc}. The M-type dwarfs are shown in Figure G.1 together with a few of the \citet{kesseli2017} spectra, illustrating the good match between our observations and the templates. All classified M dwarfs have ALFOSC spectra compatible with solar metallicity in the interval [Fe/H] = [$-1$, $+$0.5] dex. Given the low-resolution nature of our data, we could not obtain a more restrictive constraint on metallicity. This metallicity determination is also compatible with the targets' location in color-magnitude diagrams using the {\it Gaia} trigonometric parallax, where they appear on top of the main sequence of early-M type stars and clearly deviate from the locii of metal-depleted M dwarfs ([Fe/H] $\le -1$ dex). TOI-2603 (M1.0$\pm$0.5) shows H$\alpha$ in emission with a pesudo-equivalent width pEW = $-$1.6\,$\pm$\,0.2 \AA, which may hint at stellar variability. At the resolution of our data, no other target shows significant H$\alpha$ emission. The spectroscopically derived classification is consistent within one subtype with the spectral types derived photometrically from optical and near-infrared colors (e.g. \citealt{cifuentes2020}).

We generated cross-correlation functions (CCFs) of the spectra using observed standard stars of the same spectral type, aiming to detect only a single CCF peak per star. This approach enhances confidence in the confirmed planets by limiting the possibility of blended eclipsing binaries causing the transits. However, the resolution of the ALFOSC data is too low (16 Angstroms, ~140 km/s/pix) for this purpose. Binaries cannot be detected if their velocity difference at the time of observation is lower than this value. With these amplitudes, only compact companions might be visible. As expected, most CCF peaks had FWHMs ranging from 120 to 140 km/s, and no double peaks were detected.

\subsection{MuSCAT2 photometry}
Ground-based observations for all of the candidates were performed by the instrument 
Multicolour Simultaneous Camera for studying Atmospheres of Transiting exoplanets (MuSCAT2); \citet{narita2019}) mounted on the Telescopio \textit{Carlos Sánchez} (TCS) at the Observatorio del Teide (OT), Spain. This instrument is a multi-colour imager able to perform simultaneous photometry in four photometric bands (\textit{g}, \textit{r}, \textit{i}, \textit{$z_s$}) using four separate CCDs. Each channel has independent exposure times. Usually, when observing M dwarfs, the g filter's CCD is set to have shorter (<15s) exposure times and this channel’s images are normally used to auto-guide the instrument.

A specialised pipeline \citep{parviainen2020} was used to perform the data reduction and extract the aperture photometry over a given number of comparison stars and aperture sizes. We used the optimal combination of comparison stars and apertures during the transit to produce the relative light curves.

\subsection{LCOGT light curve follow-up} 

We observed one and two full predicted transit windows of TOI-2768.01 and TOI-2274.01, respectively, using the Las Cumbres Observatory Global Telescope \citep[LCOGT;][]{brown2013} 1.0\,m network nodes. We also observed a full predicted transit window of TOI-1883.01 and TOI-2274.01, using the LCOGT 2\,m Faulkes Telescope North at Haleakala Observatory on Maui, Hawai'i. The details of each observation are provided in Table E.1. We used the {\tt TESS Transit Finder}, which is a customised version of the {\tt Tapir} software package \citep{jensen2013} to schedule our transit observations. The 1\,m telescopes are equipped with $4096\times4096$ SINISTRO cameras, with an image scale of $0\farcs389$ per pixel, resulting in a $26\arcmin\times26\arcmin$ field of view. The 2\,m telescope is equipped with the MuSCAT3 multi-band imager \citep{narita2020}. The images were calibrated by the standard LCOGT {\tt BANZAI} pipeline \citep{mccully2018} and differential photometric data were extracted using {\tt AstroImageJ} \citep{collins2017}.

\subsection{High-resolution imaging}

NIRC2 installed at the 10\,m Keck2 telescope observed TOI-1883.01 in the $K$ band, TOI-2274.01 in the $K_{cont}$ band, and TOI-4838.01 in the \textit{Br-gamma} band on 24.02.2021, 23.06.2022, and 27.12.2017, respectively. The data are shown in F.1.

We observed TOI-1883 and TOI-2274 using the ShARCS camera on the Shane 3\,m telescope at Lick Observatory \citep{kupke2012, gavel2014, mcgurk2014}. 
Observations were taken with the Shane adaptive optics system in natural guide star mode to search for nearby, unresolved stellar companions. For TOI-1883, we collected a single sequence of observations using a $Ks$ filter ($\lambda_0 = 2.150$ $\mu$m, $\Delta \lambda = 0.320$ $\mu$m). For TOI-2274, we collected observation sequences using both a $Ks$ filter and a $J$ filter ($\lambda_0 = 1.238$ $\mu$m, $\Delta \lambda = 0.271$ $\mu$m). We reduced the data using the publicly available \texttt{SImMER} pipeline \citep{savel2020, savel2022}.\footnote{\url{https://github.com/arjunsavel/SImMER}} Our reduced images and corresponding contrast curves are shown in Appendix F.

TOI-2274.01, TOI-4438.01, and TOI-5319.01 were observed with the SPeckle Polarimeter (SPP; \citep{safonov2017}) in $Ic$ band on 01.05.2021, 08.11.2022, 30.11.2022, respectively (see Figure F.3). This is a facility instrument of the 2.5\,m telescope at the Caucasian Observatory of Sternberg Astronomical Institute (SAI) of Lomonosov Moscow State University. For the observations of TOI-2274, we used Electron Multiplying CCD Andor iXon 897 as a detector, while for observations of TOI-4438 and TOI-5319, we used CMOS Hamamatsu ORCA-quest. For both cases, the pixel scale is 20.6 mas/pixel, and the angular resolution is 89 mas; the field of view is 5"x5" centred on the star. The power spectrum was estimated from 4000 frames with 30 ms exposure. The atmospheric dispersion compensator was employed.  We did not detect any stellar companions brighter than $\Delta mag =3.0$ and 5.1 at 0.2" and 1.0" for TOI2274, $\Delta mag =3.3$ and 5.8 at 0.2" and 1.0" for TOI4438, $\Delta mag =3.5$ and 7.0 at 0.2" and 1.0" for TOI5319.

HRCam has observed TOI-2768.01, TOI-2781.01, and TOI-5486.01 in the \textit{I} band on 20.11.2021, 01.10.2021, and 10.06.2022, respectively. The data are shown in Figure F.4.
Finally, PHARO at the 5\,m Palomar Telescope  observed TOI-4438.01 on 19.05.2022 in the \textit{Br-gamma} band, as shown in Figure F.5.

\section{Stellar parameters}
\label{sec: Stellar_Parameters}

The stellar parameters for the stars hosting our candidates were obtained through the Exoplanet Follow-up Observation Program \citep{ExoFOP} website, which is operated by the California Institute of Technology, under contract with the National Aeronautics and Space Administration under the Exoplanet Exploration Program. The ExoFOP website is designed to expand the resources among collaborators in follow-up studies of exoplanet candidates \citep{sun2022}. ExoFOP-TESS works as a repository for the segment of the data that is gathered by the community on TESS planet candidates by permitting the exchange and display of this information and more, as mentioned in \cite{ExoFOP}. All of the parameters are displayed in both Tables C.1 and C.2.  

\section{Methods}
\label{sec: Methods}

\subsection{Multi-colour planet candidate characterisation and validation}

We analysed the TESS and ground-based light curves following the multi-colour planet validation methodology described in \cite{parviainen2019, parviainen2020, parviainen2021, parviainen2024}. 
As detailed in \citep{parviainen2024} and \citep{parviainen2019}, multi-colour planet candidate validation relies on estimating the maximum radius for the planet candidate when accounting for third-light contamination from possible unresolved stars. If this upper radius limit is below the theoretical radius limit of a brown dwarf ($\sim 0.8\,\rjup$, \citealt{burrows2011}), the candidate can be securely treated as a planet.

Contamination from unresolved sources inside a photometric aperture dilutes a transit signal, making a transit with a \enquote{true} depth, $\Delta F_\mathrm{true}$, appear to have an \enquote{apparent} depth of
\begin{equation}
\Delta F_\mathrm{app} = (1-c) \Delta F_\mathrm{true},
\end{equation}
where $c$ is the contamination, $c = F_\mathrm{c} / (F_\mathrm{c} + F_\mathrm{h})$, $F_\mathrm{c}$ is the flux from the contaminants, and $F_\mathrm{h}$ is the flux from the candidate host star. The apparent planet-star radius ratio estimated directly from the photometry, $K_\mathrm{app}$, is related to the apparent transit depth as $K_\mathrm{app} \sim \sqrt{\Delta F_\mathrm{app}}$.

Third-light contamination may lead to both wavelength-independent and wavelength-dependent changes in the shapes of the transits. First, the fact that the contaminated object's true radius ratio is different from the apparent one leads to a wavelength-independent discrepancy in the transit shape expected based on the transit depth. In theory, incorporating contamination into a transit light curve model allows for it to be estimated directly from a single-passband transit light curve. Unfortunately, however, for objects up-to Jupiter radii around sun-like host stars, this shape discrepancy is degenerate with the effects from stellar limb darkening and the object's orbital impact parameter. Furthermore, its estimation from a transit light curve requires a very high photometric precision. For M-dwarf host this situation is better, since the radius ratio of an object with a radius of $0.8\,\rjup$ spans from 0.2 (M3) to 0.8 (M8), while Neptune- and Earth-sized planets span a radius ratios of 0.08-0.35 and 0.02-0.09, respectively. The shape differences between these radius ratios and the $0.8\,\rjup$ limit are significant enough that a Jupiter-sized object (or larger) orbiting an M dwarf can be identified directly from a single-passband light curve if modelled with a transit light curve that incorporates contamination. However, the photometric quality still needs to be generally higher than what TESS can achieve due to the faintess of M dwarfs and, thus, ground-based follow-up observations are required.

In addition to the relatively weak wavelength-independent contamination effect, third-light contamination also leads to a stronger wavelength-dependent effect affecting the transit depths. This is because the level of contamination in a specific wavelength depends on the spectral types of the host star and the contaminating stars. This wavelength dependency allows us to estimate the level of contamination by combining multi-colour transit observations with a transit model that incorporates a physical contamination model. The physical contamination model is parameterised by the effective temperatures of both the host star of the planet candidate and any contaminating stars, along with a contamination factor in a designated reference passband. It computes passband-integrated contamination factors using theoretical stellar spectra from \citet{husser2013}. These calculated contamination factors are then applied to dilute the transit light curve models. The light curve model with physical contamination can then be used in a Bayesian parameter estimation framework \citep{parviainen2017}. Thus,  we are able to obtain robust estimates for the planet candidate's orbital and geometric parameters. In particular, by marginalising over the host- and comparison star temperatures, as well as the contamination levels supported by the photometry, we can derive a robust estimate of the planet-star radius ratio.

\begin{figure*}[hpbt]
    \centering
    \includegraphics[width=\textwidth]{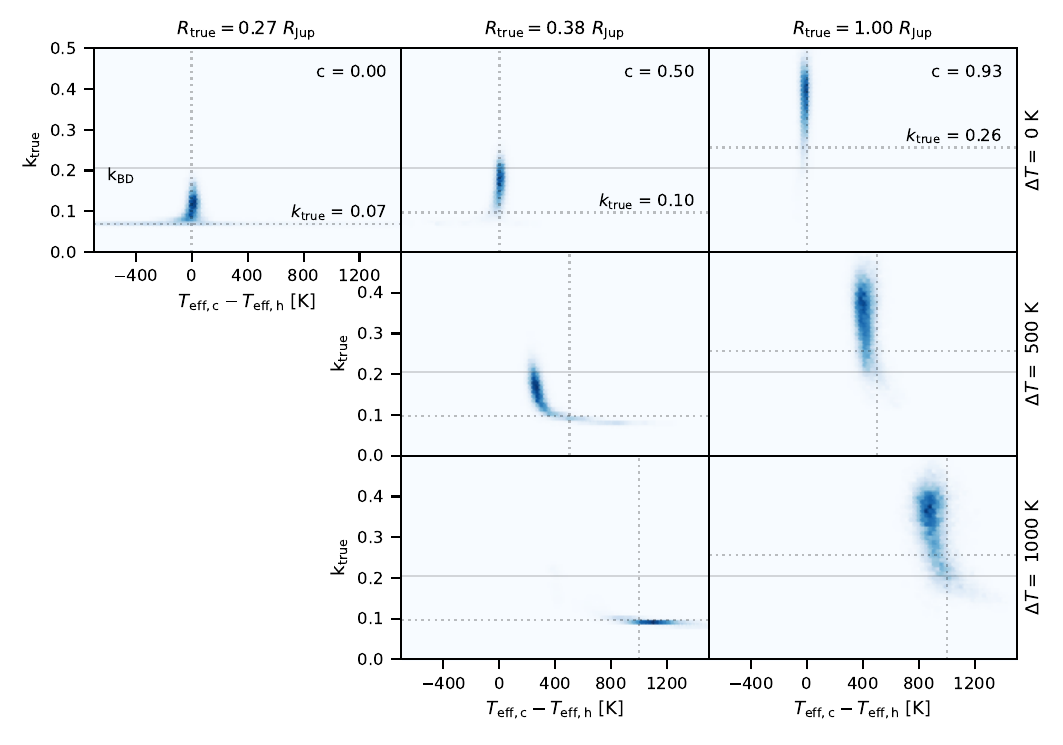}
    \caption{Multi-colour validation example using simulated light curves. The figure shows joint posterior distribution of the true radius ratios and the host-contaminant effective temperature difference for seven scenarios. The simulations assume an M dwarf host star with $T_\mathrm{eff} = 3600~K$ and $R_\star = 0.4\,R_\odot$, a transiting object with an apparent radius of $3\,R_\oplus$, three levels of contamination (0, 0.5, 0.93), and three host-contaminant temperature differences (0~K, 500~K, 1000~K). The $c=0$ scenario corresponds to an uncontaminated object, while $c=0.5$ corresponds to a scenario that would be expected if the contaminating star is similar to the host star (i.e. maximum amount of contamination that can be expected without significant colour difference), and the $c=0.93$ scenario corresponds to a contaminated Jupiter-sized object. The dotted vertical lines show the true $\Delta \teff$, the dotted horizontal line shows the true radius ratio, and the solid horizontal line shows the radius ratio limit corresponding to our chosen $0.8\,\rjup$ brown dwarf radius limit.}
    \label{fig:posterior_example}
\end{figure*}

Figure~\ref{fig:posterior_example} shows the joint posterior distributions for the true radius ratio and host-contaminant temperature difference for a set of simulated (realistic) contamination scenarios.\!\footnote{The code reproducing the simulations, including visualisations of the simulated light curves, is available in Jupyter notebooks \url{https://github.com/hpparvi/PyTransit/blob/master/notebooks/contamination/2024_m_dwarf_host_example_simulations.ipynb} and \url{https://github.com/hpparvi/PyTransit/blob/master/notebooks/contamination/2024_m_dwarf_host_example_plots.ipynb}} We assumed an M~dwarf host with $T_\mathrm{eff} = 3600~K$ and $R_\star = 0.4\,R_\odot$,  and a transiting object with an apparent radius of $3\,R_\oplus$. We repeated the simulations for three contamination scenarios (columns, $c\in\{0.0, 0.5, 0.93\}$) and three effective temperature differences (rows, $\Delta T_\mathrm{eff} \in \{0~K, 500~K, 1000~K\}$). For the $\Delta\teff = 0$ scenarios, the \ktrue posteriors strongly reject any contamination from a star (or stars) of different spectral type than the host; the posterior near the $\Delta\teff = 0$ line is constrained by the colour-independent effect contamination has on the transit shape. For the scenarios with $\Delta\teff \neq 0~K$, the true solution is covered by the tail of the posteriors (the match would improve with a higher S/N), and increasing the colour difference leads to an improved \ktrue estimate; we note that this is assuming a single contaminant, as the situation would be more complex if we were to allow for several contaminating stars.

The multi-colour analysis presented here is computed by a specialised pipeline that uses a transit model incorporating physical contamination model to model both the wavelength-dependent and independent contamination effects (implemented in \texttt{PyTRANSIT} \citep{parviainen2015pytransit}). Furthermore, the pipeline uses \texttt{LDTk} \citep{parviainen2015ldtk} for limb darkening estimations and \texttt{emcee} \citep{foreman2013} to perform the MCMC sampling.

For each target, we applied the multi-colour analysis jointly to all the available data sets, including the MuSCAT2 and TESS light curves, as well as the LCOGT light curves, if available. The stellar parameters used for our calculations are shown in Tables C.1 and C.2.
The pipeline used to perform this multi-colour analysis calculates the system parameters, the apparent radius ratio ($K_{app}$) distribution, contamination-corrected true radius ratio ($K_{true}$) distribution, the effective temperature ($T_{eff}$) of the companion, and the impact parameter (b). All these parameters allow us to determine the true nature of a transit signal and decide whether the stellar host companion is a planetary object.

\begin{figure*}[hpbt]
    \centering
    \includegraphics[width=\textwidth]{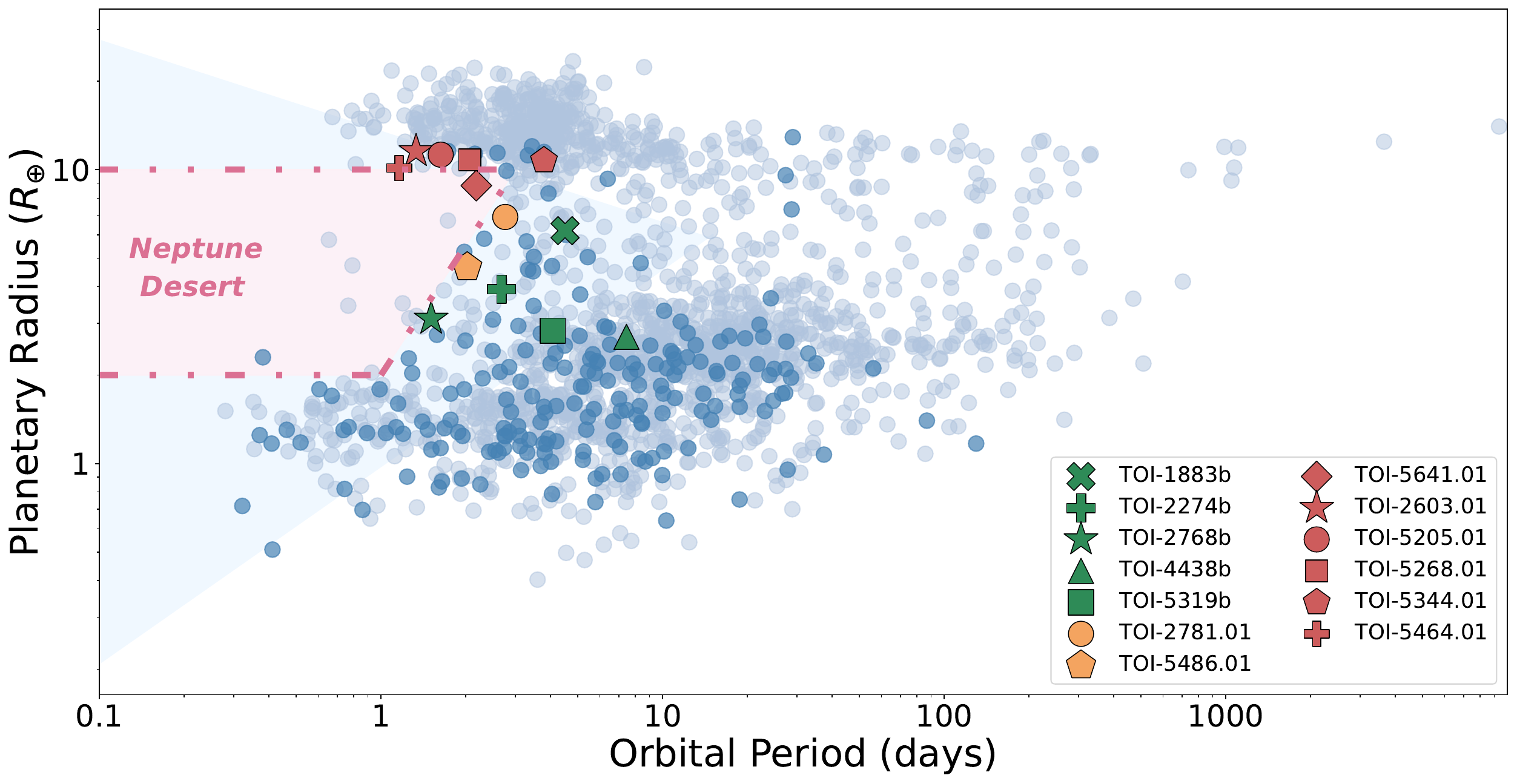}
    \caption{Location of the validated and non-validated planet candidates in the period-radius diagram. The red polygons show the non-validated candidates, the dark-green polygons show the validated planets, and the orange polygons show the candidates which validation status is only suggestive. The grey circles show the population of all validated planets with radius uncertainties below $10\%$, and the blue circles highlight those planets orbiting M-type stars. The blue-shaded region shows the Neptune desert as defined by \cite{mazeh2016}. However, in recent years, this region has become more populated, and the boundaries of the desert do not match the typical definitions any longer. Thus, we suggest the pink-shaded area, where the scarcity of planets is noteworthy, as an updated definition of the Neptune desert based on the current population of known planets.}
    \label{fig:demography}
\end{figure*}

\subsection{False alarm probability estimation}


We multiplied each $K_{true}$ distribution by a Gaussian distribution corresponding to the host star radius to obtain the posterior distribution of the absolute true radius of each transiting object, accounting for potential third-light contamination effects and the uncertainties in the stellar radii. The absolute true radius distribution offer a natural way to estimate the false alarm probability (FAP) of a planetary candidate as the percentage of samples that fulfills $R_p\ > 0.8\ R_J$. In some cases, there are no samples above the radius threshold, for which we set a minimum FAP as if there was at least one sample. Tables A.1 and A.2 report the FAPs calculated with this method for all candidates with median true radius below the threshold. We note that a higher radius does not rule out the planetary nature of the transiting object, as there are more than 800 confirmed planets with $R_p\ > 0.8\ R_J$ to date. Our FAPs are very conservative upper limits, as they include giant planets as false positives, due to their radii being consistent with those of brown dwarfs. Furthermore, the upper tail of the true radius distributions correspond to scenarios with significant contamination from a blended source of the same spectral type or similar to the host star. Our complementary observations with high-resolution imaging indicate that such scenarios are unlikely for the targets considered in this work.



\section{Results}
\label{sec: Results}

For order and clarity, we show the individual MuSCAT2 and TESS light curves, together with the MuSCAT3 and LCO-SINISTRO light curves (when available), for each target in Appendix B.1 to B.13. We also show, in Appendices D.1 and D.2, the results of the multi-colour analysis for each target displaying the joint posterior distributions of true radius ratio, $K_{true}$, and the difference in effective temperature between the host and a possible contaminant ($\Delta T_{eff}$), the apparent radius ratio, $K_{app}$, the impact parameter, and the stellar density in $g/cm^{3}$ for each candidate.

The derived planet parameters for each object in our sample are given in Tables A.1 and A.2. For reference, we show all the planetary candidates in a radius-period diagram in Fig.~\ref{fig:demography}, where we compare them with the rest of the known planets with radius uncertainties below 10\%. 

\begin{table}[hpbt]
\caption{Prior probability distributions of the fitted parameters.}
\centering
\begin{tabular}{lcc}
  \hline
  \noalign{\smallskip}
  Description &  Parameter & Value \\
  \hline
  \noalign{\smallskip}
  Global parameters \\
  \hline
  \noalign{\smallskip}
  \noalign{\smallskip}

  Orbital period & P (days) & $\mathcal{N}^{(a)}$ \\
  Stellar density & $\rho_*$ (g/cm$^{-3}$) & $\mathcal{U}$(1,35) \\  
  Impact parameter & b & $\mathcal{U}$(0,1) \\
  Zero epoch & T$_0$ & $\mathcal{N}^{(a)}$ \\  
  Robust area ratio & k$^2_{true}$ & $\mathcal{U}$(1,0) \\ 
  Host temperature & T$_h$ (K) & $\mathcal{N}^{(a)}$ \\  
  Cont. temperature & T$_c$ (K) & $\mathcal{U}$(1200, 7000) \\
  
  \noalign{\smallskip}
  \noalign{\smallskip}
  Passband-dependent \\
  parameters \\
  \hline
  \noalign{\smallskip}
  \noalign{\smallskip}
  Limb darkening & q & $\mathcal{N}^{(b)}$ \\

\noalign{\smallskip}
  \noalign{\smallskip}
  Light-curve-dependent \\
  parameters \\
  \hline
  \noalign{\smallskip}
  \noalign{\smallskip}
  Log$_{10}$ white noise & log$_{10}\ \sigma$ & $\mathcal{U}$(-4,0)$^{(c)}$ \\
  \noalign{\smallskip}
  \noalign{\smallskip}
  \hline
\end{tabular}
\vspace{0.5cm}
\tablefoot{Transit light curve model parameters and priors. Individual target posteriors are shown in Tables A.1 and A.2. The global parameters are independent of the passband or light curve. The passband-dependent parameters are specified for each passband, and the light-curve-dependent parameters are repeated for each individual light curve. $\mathcal{N}$ stands for a normal prior, and $\mathcal{U}$ (a, b) stands for a uniform distribution from a to b.\\
\tablefoottext{a}{Centered on the ExoFOP values. Prior $\sigma$ equals three times the nominal error bars. For the uniform distributions, the whole interval equals six times the nominal error bars.}\\
\tablefoottext{b}{Limb darkening coefficients correspond to the transformed power-2 limb darkening law coefficients \citep{hestroffer1997,morello2017,short2019} and have normal priors calculated using \texttt{LDTK}.}\\
\tablefoottext{c}{Average log$_{10}$ white noise parameters for each light curve have uninformative uniform priors.}
}\\

\label{tab:priors}
\end{table}
        
\subsection{Validated planets}

The validation process corroborates that the observed signal is due to a planetary transit rather than some other astrophysical phenomenon.
Therefore, candidates that show a low degree of contamination and whose radii are below the typical radius of a brown dwarf (with 99\% probability, FAP$<$1\%) were validated as planets.
We adopted the conservative criterion of considering the radius limit above which an object can be a brown dwarf at $0.8 R_J$ \citep{burrows2011}, which is a discriminatory key parameter for target validation. Thus,  candidates with a radius close to this limit (even though they might exhibit a planetary nature) cannot be validated as planets only based on size criteria.
In addition, the impact of contamination in the host source flux was considered by utilising the comparison between $K_{app}$ and $K_{true}$. Our analysis suggests that contamination seems to be negligible for the seven objects discussed in this subsection.

TOI-1883b, TOI-2274b, TOI-2768b, TOI-4438b, and TOI-5319b were validated as planets with sizes ranging from sub-Jovian to sub-Neptune and super-Earth. TOI-2781.01 and TOI-5486.01 are most likely exoplanets, although they would need at least an additional multi-band observation to be formally validated. Therefore, given the currently available data, we can only claim strong evidence for their planetary nature. For comparison, TOI-5486.01 has a similar size and low impact parameter to the sub-Neptune TOI-5319b, with no signs of colour contamination or V-shaped light-curves. The case of TOI-2781.01 is less obvious, as it would be the largest validated exoplanet of our sample, but similar to TOI-1883b.
As shown in the figures of Appendix B, there is no significant chromaticity and no signs of strong contamination in the MUSCAT2 light curves. TOI-2274 is located close (2.48\arcsec) to a faint ($\Delta T = 4.61$) star (TIC 1550481885); in theory, it could also be the host to the object causing the transit signal. However, a solution where the faint background star is be a binary diluted by TOI-2274 would lead to very strong colour signal because the contamination would be $\approx98\%$. The solution where TOI-2274 is the host leads to contamination of $\approx2\%$ from the known background star, which does not affect the radius ratio estimate in any significant way. TOI-2768b, TOI-2781.01, and TOI-5486.01 lie near the boundaries of the Neptune Desert region in the radius-period diagram.

For validated planets with no signs of blends, we chose to report the apparent radius as, in absence of blends, the true radius distribution is systematically biased to higher radii \citep{parviainen2019}.
Some of the candidates have also been validated by other groups. That is the case for TOI-5205.01, which was validated by \cite{kanodia2023}. Although we could not validate the planet (Section \ref{sec: non-validated}), it is a good example of the limits of the method used. TOI-5344.01 was recently also validated by both \cite{hartman2023} and \cite{Han_2024}. Finally, TOI-4438.01 was recently validated by \cite{Goffo2024}.


\subsection{Non-validated planet candidates}
\label{sec: non-validated}

A false positive can occur when a star's light is dimmed by something other than a planet passing in front of it. Indications of this can be observed in the light curves, where there may be patterns that do not match the expected signature of a transiting planet. One such pattern is the significant presence of a colour dependence in the transit depth in different spectral bins. This is the case for TOI-2603.01 and TOI-5268.01, where significant chromaticity is seen in the MuSCAT2 light curves. Therefore, additional investigation and analysis will be necessary to establish the nature of these two candidates.

Exoplanet candidates with radii exceeding the upper limit of $0.8 R_J$ (including uncertainties) can not be validated using only multi-colour photometry. Thus, we were unable to validate TOI-5205.01, TOI-5344.01, TOI-5464.01, and TOI-5641.01 due to their high true radius values, enough to be above the lower limit for a brown dwarf (although still in the planetary regime also). TOI-5641.01 has an apparent radius smaller than, but perhaps too close to the critical radius for being robustly validated with a small amount of photometric observations. These three candidates do not show a significant colour signature and all of them lie within the gas giant region in the radius-period diagram, some bordering the Neptune desert region.

\section{Discussion and conclusions}
\label{sec: Discussion}

To bolster the interpretation of our results in the context of planetary demographics, we located our candidates in the period-radius diagram and compared them with the population of all known exoplanets with radius uncertainties below 10\% (Fig.~\ref{fig:demography}). In the figure, we highlight the M-dwarf hosted population ($T_{eff}<4000K$) as our candidates are orbiting M-type host stars. In green, we marked the candidates that we were able to validate and in red the candidates that we were unable to validate. 

We also highlighted in blue the region of the Neptune desert matching the definition of \cite{mazeh2016}. Over the years, however, new planet discoveries have shown that the extension of the Neptune desert proposed by \cite{mazeh2016} does not match the exoplanet distribution any more. Here, we propose a new empirical definition of the Neptune desert, which simply encompass the radius region between 2 and 10 $R_\oplus$ and extends to periods from 1 to 3 days for these two radii, respectively. The new proposed region is marked in pink.

The number of new targets near or within the Neptune desert validated in our manuscript is too small to provide new statistical population insights. However, validated planets around and inside this desert are extremely interesting targets to be further explored for mass and bulk density determinations and, ultimately, for atmospheric characterisation studies as well. These subsequent studies (given that composition can be traced to formation and atmospheric evolution processes) will ultimately be able to shed light into the physical nature of the Neptune desert. This would offer valuable insights into the formation and composition of these exoplanets and their atmospheres.

In particular, the planetary nature of TOI-2781.01 and TOI-5486.01, which lie exactly within the Neptune desert boundary, has not yet been validated. However, our results show compelling evidence that these two candidates may soon be validated as planets. There is no evidence of color contamination in the light curves of either candidate. The FAP values for both candidates range between 1\% to 50\%. By adding further observations to the analysis of these planets, we could potentially lower the FAP values below the 1\% threshold. The validated candidate with the largest radius is TOI-1883.01, with $R_p=5.65\ R_{\oplus}$. Since the true radius value of TOI-5486.01 is smaller than that of TOI-1883b, we can confidently state that with one additional observation, the FAP value for TOI-5486.01 would fall below the 1\% limit. The apparent radius of TOI-2781.01 is larger than that of TOI-1883b. However, due to the larger errors in the radius and $K_{app}$ compared to those of TOI-1883b, we can assess that additional observations would be required to validate it. TOI-2768 b is validated in this work and also lies in the same neptune desert boundary, with a smaller radius.

It is also noteworthy that many of our planet candidates have large radii, but these type of planets are uncommon around M dwarf hosts \citep{Morales2019}. While we did not validate planets with apparent radii larger than of $0.8 R_J$, TOI-5205.01, TOI-5344.01, and TOI-5464.01 show no evidence of chromaticity in their light curves; thus, they are candidates for the enlargement of the rare population of giant planets around M dwarfs. While the host stars are faint, the expected radial velocity semi-amplitudes are large and can be measured with state-of-the-art instruments such as ESPRESSO or MAROON-X (among others). This would allow for further orbital and atmospheric characterisations. Within the group of our validated planets, TOI-1883b stands out as a very interesting candidate, with a warm equilibrium temperature of 524 K.


In summary, we have discussed and validated the planetary nature of at least five new planet candidates and their properties. We have also evaluated the degree of contamination in the flux of each object through the analysis of the true and apparent radius ratio. TOI-2781.01 and TOI-5486.01 remain in the suggestive evidence category, awaiting more observations to be thoroughly validated. We were unable to validate the other six planets in our sample due to chromaticity in the light curves and/or a retrieved true radius value that overlaps with the brown dwarf regime.


\begin{acknowledgements}


Funding for the TESS mission is provided by NASA's Science Mission Directorate.

This article is based on observations made with the MuSCAT2
instrument, developed by ABC, at Telescopio Carlos Sánchez operated on the island of Tenerife by the IAC in the Spanish Observatorio del Teide.

This work makes use of observations from the LCOGT network. Part of the LCOGT telescope time was granted by NOIRLab through the Mid-Scale Innovations Program (MSIP). MSIP is funded by NSF.

This paper is based on observations made with the MuSCAT3 instrument, developed by the Astrobiology Center and under financial supports by JSPS KAKENHI (JP18H05439) and JST PRESTO (JPMJPR1775), at Faulkes Telescope North on Maui, HI, operated by the Las Cumbres Observatory.

Based on observations made with the Nordic Optical Telescope, owned in collaboration by the University of Turku and Aarhus University, and operated jointly by Aarhus University, the University of Turku and the University of Oslo, representing Denmark, Finland and Norway, the University of Iceland and Stockholm University at the Observatorio del Roque de los Muchachos, La Palma, Spain, of the Instituto de Astrofisica de Canarias.

The data presented here were obtained in part with ALFOSC, which is provided by the Instituto de Astrofisica de Andalucia (IAA) under a joint agreement with the University of Copenhagen and NOT.

This research has made use of the Exoplanet Follow-up Observation Program (ExoFOP; DOI: 10.26134/ExoFOP5) website, which is operated by the California Institute of Technology, under contract with the National Aeronautics and Space Administration under the Exoplanet Exploration Program.

We acknowledge the use of public TESS data from pipelines at the TESS Science Office and at the TESS Science Processing Operations Center. Resources supporting this work were provided by the NASA High-End Computing (HEC) Program through the NASA Advanced Supercomputing (NAS) Division at Ames Research Center for the production of the SPOC data products.

This work has made use of data from the European Space Agency (ESA) mission {\it Gaia} (\url{https://www.cosmos.esa.int/gaia}), processed by the {\it Gaia} Data Processing and Analysis Consortium (DPAC, \url{https://www.cosmos.esa.int/web/gaia/dpac/consortium}). Funding for the DPAC has been provided by national institutions, in particular the institutions participating in the {\it Gaia} Multilateral Agreement.

Funding from University of La Laguna and the Spanish Ministry of Universities is acknowledged.
HP acknowledges support by the Spanish Ministry of Science and Innovation with the Ramon y Cajal fellowship number RYC2021-031798-I.

A. P-T. acknowledges financial support from the Severo Ochoa grant CEX2021-001131-S funded by MCIN/AEI/ 10.13039/501100011033.

EP acknowledges funding from the Spanish Ministry of Economics and Competitiveness through project PID2021-125627OB-C32

This work is partly supported by JSPS KAKENHI Grant Number JP18H05439 and JST CREST Grant Number JPMJCR1761.

E. E-B. acknowledges financial support from the European Union and the State Agency of Investigation of the Spanish Ministry of Science and Innovation (MICINN) under the grant PRE2020-093107 of the Pre-Doc Program for the Training of Doctors (FPI-SO) through FSE funds.

J.~K. gratefully acknowledges the support of the Swedish National Space Agency (SNSA; DNR 2020-00104) and of the Swedish Research Council  (VR: Etableringsbidrag 2017-04945)

G.~M. has received fundings from the Ariel Postdoctoral Fellowship program of the Swedish National Space Agency (SNSA), the Severo Ochoa grant CEX2021-001131-S and the Ram\'on y Cajal grant RYC2022-037854-I funded by MCIN/AEI/ 10.13039/501100011033 and FSE+.

M.S. acknowledge the support of the Italian National Institute of Astrophysics (INAF) through the project 'The HOT-ATMOS Project: characterising the atmospheres of hot giant planets as a key to understand the exoplanet diversity' (1.05.01.85.04).

This work is partly supported by JSPS KAKENHI Grant Number JP21K20376.

This work is partly supported by Astrobiology Center SATELLITE Research project AB022006

This work is partly supported by JSPS KAKENHI Grant Number JP21K13955.

D.V.C. and S.G.Z. acknowledge the support of Ministry of Science and Higher Educationof the Russian Federation under the grant 075-15-2020-780 (N13.1902.21.0039).

R.L. acknowledges funding from University of La Laguna through the Margarita Salas Fellowship from the Spanish Ministry of Universities ref. UNI/551/2021-May 26, and under the EU Next Generation funds.

The data presented here were obtained in part with ALFOSC, which is provided by the Instituto de Astrofisica de Andalucia (IAA) under a joint agreement with the University of Copenhagen and NOT.
MRZO acknowledges financial support from project PID2019-109522GB-C51 funded by the Spanish Ministry of Science and Innovation. 
      
\end{acknowledgements}

%
%

\bibliographystyle{aa}
\bibliography{Biblio}

\end{document}